# Enhancing Cluster Scheduling in HPC: A Continuous Transfer Learning for Real-Time Optimization

Leszek Sliwko
*School of Computer Science and Engineering*
University of Westminster
London, United Kingdom
ORCID: 0000-0002-1927-8710

Jolanta Mizera-Pietraszko
*Department of Computer Science*
Opole University of Technology
Opole, Poland
ORCID: 0000-0002-2298-5037

*Abstract*—This study presents a machine learning-assisted approach to optimize task scheduling in cluster systems, focusing on node-affinity constraints. Traditional schedulers like Kubernetes struggle with real-time adaptability, whereas the proposed continuous transfer learning model evolves dynamically during operations, minimizing retraining needs. Evaluated on Google Cluster Data, the model achieves over 99% accuracy, reducing computational overhead and improving scheduling latency for constrained tasks. This scalable solution enables real-time optimization, advancing machine learning integration in cluster management and paving the way for future adaptive scheduling strategies.

## I. INTRODUCTION

In the rapidly evolving landscape of cloud computing and distributed high-performance environments, the efficient management of architectural and software resources became apparently paramount for ensuring suitable performance and minimizing latency. As long as the industry organizations increasingly rely on cluster-based architectures to orchestrate their broad areas of possible applications, the importance of effective task scheduling has come to the forefront. Over the last few years, traditional schedulers, such as Kubernetes and some more, have laid the groundwork for managing containerized workloads; however, it was found that it poses a challenge for them to adapt to the dynamic nature of real-time workloads and node-affinity constraints [35]. These limitations result in inefficient resource utilization and longer scheduling delays, which ultimately affect overall system performance, especially in high-performance systems [9][18]. In mission-critical environments, these issues can escalate, disrupting vital systems like power networks, healthcare, defense systems, and others. Thus, it is crucial to implement robust scheduling strategies that can manage high and dynamic workloads effectively.

To address all these kinds of challenges, nowadays, the integration of Machine Learning (ML) techniques into scheduling systems has emerged as a promising solution [7][8][12]. By leveraging historical data and learning algorithms in real time, these methods can enhance the decision-making processes associated with task scheduling. Nevertheless, the existing ML models usually require significant retraining periods to adapt to the constantly changing workload patterns, which can render them impractical in high-velocity scenarios, in which both the resources and demands fluctuate frequently.

This study aims to bridge this gap by proposing a continuous learning method for cluster scheduling systems that utilizes a Transfer Learning (TL) model. Unlike conventional approaches proposed in the state-of-the-art, the presented method dynamically evolves during operation, minimizing the need for retraining all the data while maintaining high accuracy of the outcome and low computational overhead thanks to the high-complexity algorithms used. Through a thorough evaluation based on Google Cluster Data traces, it is demonstrated that the model not only achieves remarkable accuracy exceeding 99% but also significantly reduces scheduling latency for tasks with restrictive node-affinity constraints.

The extensible nature of presented novel solution paves the way for real-time optimization and increased scalability, enhancing the capabilities of the cluster management systems simultaneously. As this paper delves deeper into the research, it will explore the implications of ML integration in scheduling strategies and its potential to pave the way for developing adaptive scheduling methodologies that can meet the needs of diverse and complex workloads. Additionally, this approach can contribute to improving resource utilization and, thus, reducing operational costs. Specifically, the objective of this study is to highlight the transformative potential of continuous learning models in enabling more intelligent and responsive cluster scheduling for high-performance computing.

The Continuous Transfer Learning Method (CTLM) for cluster scheduling systems is defined as a ML approach designed to optimize task scheduling in distributed environments while accounting for node-affinity constraints. The investigated model involves utilizing some pre-training that can adapt to new upcoming data in dynamical settings while evolving the workload circumstances in real time. It's due to the fact that CTLM continuously learns from ongoing operations within the cluster, thereby enabling the system to refine scheduling decisions dynamically. Such that the method presented in this paper leverages historical scheduling data, ensures minimal retraining, and focuses on maintaining high accuracy amidst changing conditions and a variety of scenarios.

This paper has the following **key contributions**:

- It introduces a novel approach to cluster scheduling, known as the continuous learning method with node-affinity constraints, in high-performance computing,
- It provides an overview of the present body of knowledge on ML-based cluster scheduling paradigms like RL, DL, and some more,
- It demonstrates the potential of continuous cluster scheduling in high-performance scenarios and highlights the applicability of optimization algorithms in cloud computing environments,
- It discusses Google Cluster Data (GCD) traces logical operators for managing the task constrains,
- It presents a new approach to TL for cluster scheduling in a high-performance variety of computing scenarios.

The remainder of this paper is organized as follows: Section II discusses a state-of-the-art of optimization approaches to cluster workload allocation. Section III presents a step-by-step simulation on cluster scheduling based on Google cluster data-traces logical operators. Section IV describes in detail the dynamically growing model in high-performance scenarios supported by optimization algorithms. Section V demonstrates the advantages of the evaluation optimizers implemented and discusses the findings. Finally, Section VI concludes the research providing the future work plan.

## II. RELATED WORKS

The optimization of cluster workload allocation has been a longstanding focus in distributed computing, with systems like SLURM, Kubernetes, and Google's Borg addressing scalability and resource management challenges. However, issues such as node-affinity constraints and dynamic workload adjustments remain unresolved. To tackle these, researchers have explored ML techniques, including Reinforcement Learning (RL) and Neural Networks (NN), to enhance adaptability and efficiency in scheduling.

Recent advances integrate TL and deep RL to improve resource utilization and reduce computational costs. TL enables reusing pre-trained models, while RL adapts dynamically to scheduling constraints. Despite these strides, challenges persist in scaling solutions and handling heterogeneous workloads. This section highlights key contributions in scheduling systems, ML applications, and simulation frameworks, emphasizing their relevance to this study.

### A. Task Scheduling in Cluster Environments

Task scheduling in cluster environments has been a focus of extensive research for many years, leading to the development of robust and efficient systems. Here are some notable systems and their contributions to the field:

- **SLURM**: An open-source, scalable cluster management and job scheduling system for HPC, supporting resource allocation, queues, multi-node jobs, and flexible scheduling policies like fair-share and backfilling, with integration for resource monitoring [14].
- **Microsoft Apollo**: It handles high task churn, processing 100k+ requests/sec on 20k-node clusters using per-job Job Managers and local Process Nodes, prioritizing smaller tasks for immediate execution [15].
- **Alibaba Fuxi**: This system uses a unique approach by matching newly available resources to the backlog of tasks, rather than matching tasks to resources, achieving 95% memory and 91% CPU utilization, scaling to 5,000 nodes since 2009 [16].
- **Twitter Aurora**: It manages batch/real-time workloads, integrates with Apache Mesos, supports fault tolerance via rescheduling/checkpointing, and optimizes resources through task packing [17].
- **Google's Borg and Omega**: Borg, Google's computing cell scheduler, runs multiple parallel schedulers, initially using the Enhanced Parallel Virtual Machine algorithm but later adopting a hybrid fairness and best-fit model to reduce fragmentation and improve resource handling. Omega tackled scalability and head-of-line blocking with Paxos-based state storage and optimistic locking, enabling higher throughput. Many of these innovations were integrated back into Borg [18].

While these systems are proprietary, open-source alternatives such as Docker Swarm and Kubernetes are widely used for task scheduling. Kubernetes, in particular, has become a leading platform for distributed scheduling and resource orchestration due to its flexibility and extensibility.

### B. Machine Learning in Cluster Scheduling

Recent studies highlight the transformative potential of ML in cluster scheduling and load balancing, offering scalable solutions to complex scheduling challenges. Unlike traditional algorithms, ML-based approaches leverage historical data to predict resource demands and make adaptive decisions. Key advancements include:

- **Reinforcement Learning**: Multi-agent RL has been applied to task load balancing in cloud-edge environments, enabling agents to learn suitable scheduling policies and outperform conventional methods [29].
- **Deep Learning Models**: A dynamic load-balancing strategy Convolutional Neural Networks (CNN) and

Recurrent Neural Networks (RNN) optimizes task scheduling and workload distribution in cloud systems [30].

- **Proactive Container Scheduling**: A predictive control has been used for long-term load balancing by migrating long-running workloads in shared clusters [31].

These innovations are categorized NN-based scheduling (Table I), load balancing with deep learning (Table II), practical cloud and SDN deployments (Table III), and other advanced ML techniques (Table IV).

TABLE I.  NEURAL NETWORKS AND WORKFLOW SCHEDULING

| Ref. | Description |
|---|---|
| [5] | Proposes a scheduling model using NN combined with RL. The model employs an encoder to vectorize workflow characteristics, including task properties and resource availability, to estimate execution times. |
| [6] | Leverage Long Short-Term Memory (LSTM) methods for workload prediction. These models, combined with optimization techniques like Particle Swarm Intelligence, enhance dynamic workload provisioning and minimize latency. |
| [23] | Proposes an RL-based scheduler that mitigates network contention in GPU clusters. The approach dynamically adapts scheduling decisions based on contention sensitivities, leading to reductions in average and tail job completion times compared to traditional policies. |
| [31] | A model predictive control-based container scheduling strategy was introduced to proactively migrate long-running workloads for long-term load balancing in shared clusters. |
| [33] | Focus on real-time workload forecasting and resource allocation using a mix of ML algorithms like random forests, SVMs, and RL. |

TABLE II.  LOAD BALANCING WITH DEEP LEARNING MODELS

| Ref. | Description |
|---|---|
| [7] | Replaces traditional hash functions in load balancing mechanisms with deep learning models. These models are trained to uniformly map workload distributions across servers, ensuring balanced resource use. |
| [8] | Introduces a dynamic ML-based load balancer that selects the most suitable strategy based on historical workload data, showcasing the adaptability of ML methods. |
| [9] | Extends Kubernetes with ML modules, using RL to optimize scheduling decisions. It demonstrates improved cluster load distribution by dynamically selecting worker nodes with the shortest response times. |
| [30] | Developed a dynamic load balancing strategy using a deep learning model with CNNs and RNNs to optimize task scheduling and enhance cloud performance. |
| [31] | A model predictive control-based container scheduling strategy was introduced to proactively migrate long-running workloads for long-term load balancing in shared clusters. |

TABLE III.  APPLICATIONS IN CLOUD AND SDN ARCHITECTURES

| Ref. | Description |
|---|---|
| [12] | Introduces a workload prediction framework combining NN with a differential evolution algorithm for parameter optimization. |
| [13] | Employs LSTM models with evolutionary algorithms for optimizing metrics such as latency, throughput, and cost in cloud environments. |
| [29] | Introduces multi-agent RL frameworks for task load balancing in cloud-edge environments, where agents learn suitable scheduling policies through interactions, outperforming traditional algorithms. |

TABLE IV.  ADVANCED TECHNIQUES IN ML-BASED SCHEDULING

| Ref. | Description |
|---|---|
| [10] | Combines Support Vector Machines (SVM) and K-means clustering for classifying and grouping Virtual Machine resources based on predicted utilization. |
| [11] | Investigates sampling-based runtime estimation without using ML. Tasks are sampled and analyzed to predict overall job runtime properties, achieving significant reductions in average completion times. |
| [21] | Demonstrates that modern ML techniques, specifically RL, can automatically generate highly efficient scheduling policies for data processing clusters, outperforming traditional heuristics. |
| [22] | Presents a scheduler for ML workloads in clusters, leveraging deep RL techniques to improve scheduling decisions based on accumulated experience. |

## C. Transfer Learning in Neutral Networks

TL is a powerful technique in deep learning and NN, where a model trained on one task is adapted to perform a different, yet related, task. This approach leverages pre-trained models that have already learned useful features from other datasets, thereby saving time and computational resources when addressing new problems. The technique is based on the principle that the lower layers of NN capture general features (e.g., edges in images or basic linguistic patterns) that can be effectively reused for new tasks. Meanwhile, the upper layers, responsible for task-specific outputs, are fine-tuned to address the specific problem. This adaptability makes TL particularly effective in scenarios with limited labelled data, as it eliminates the need to train models from scratch – a process that often requires large datasets and significant computational power. A comprehensive survey [4] offers insights into the theoretical foundations of TL.

In NLP, models like BERT [19] and GPT [20] are pre-trained on large text corpora and fine-tuned for tasks such as sentiment analysis and text summarization, achieving state-of-the-art results. In computer vision, pre-trained models like VGG, ResNet, and EfficientNet [34] are fine-tuned for applications such as cancer detection or galaxy classification. Similarly, in speech recognition and synthesis, pre-trained models are adapted for specific languages or accents. TL also excels in fields with scarce labeled data, such as bioinformatics and materials science, enabling advances like protein structure prediction and drug-target interaction analysis. Its adaptability and efficiency have made it foundational in modern AI research and applications.

## III. CLUSTER SCHEDULING SIMULATION

Analyzing distributed applications and services without full access to computing clusters is challenging due to the unique nature of cloud workloads, which differ from traditional grid computing [24]. Publicly available cloud workload traces are scarce and often lack critical details [25], leading researchers to rely on simulations and models.

The AGOCS project [1], developed between 2015 and 2018 to create a distributed cluster orchestration system [26], highlighted the importance of realistic input data for accurate outcomes. Cloud systems' complexity requires simplifications in simulations, limiting their ability to represent realistic configurations, especially for system-critical mechanisms like task scheduling or fault handling. To address this, AGOCS used workload traces from the GCD archive [2]. These traces were parsed and replayed, simulating scheduler operations.

*A. Google Cluster Data traces*

The key elements of AGOCS include processing collections and task events, handling machine events and updates, and matching tasks to available machines based on task constraints. The logic behind this matching is the focus of this investigation. As mentioned, AGOCS was originally built upon the GCD traces from 2011, which specified four logical Constraint Operators (CO) (coded as numeric values):

- **Equal** operator: The node's attribute must match the specified constraint or remain empty if no value is specified; applies to numeric and non-numeric values.

- **Not-Equal** operator: The attribute must be absent or differ from the specified constraint; applies to numeric and non-numeric values.

- **Less-Than** operator: For numeric values, requiring the attribute to be less than the specified constraint.

- **Greater-Than** operator: For numeric values, requiring the attribute to exceed the specified constraint.

In April 2020, the GCD archive was updated with May 2019 traces [2]. These 31-day traces include new features like alloc sets, batch queuing, vertical scaling, and power utilization logs (powerdata-2019) for 57 Google data center power domains, with detailed changes outlined in [3]. Borg now uses abstract Google Compute Units instead of CPU core counts, mapping them to physical cores as needed. The 2019 traces also add task parent-child dependencies and four new logical COs:

- **Less-Than-Equal** operator: For numeric values, requiring the attribute to be ≤ the specified constraint.

- **Greater-Than-Equal** operator: For numeric values, requiring the attribute to be ≥ the specified constraint.

- **Present** operator: Ensures the attribute is defined and non-blank; applies to numeric and non-numeric values.

- **Not-Present** operator: Ensures the attribute is undefined; applies to numeric and non-numeric values.

The GCD clusterdata-2019 traces include data from eight computing cells (A–H) instead of one, with a similar cell size of 12.1k–12.6k machines (9.4k for cell A). The format shifted from 2011 CSV files to a Google BigQuery-stored dataset (~2.4 TB compressed). For this research, the data was downloaded, sorted by timestamp, and the AGOCS tool [1] was adapted to the clusterdata-2019 JSON format. However, the traces presented anomalies, including (i) inaccurate event timings, where task updates occurred before terminations (e.g., eviction, failure, completion), and (ii) tasks missing eviction or failure events, complicating task removal. To address this, AGOCS was modified to auto-correct event timings (e.g., offsetting updates after creation) and synchronize task marker removal with collection events, ensuring terminated collections deleted associated task markers.

In ML systems, dataset preparation is as crucial as model building. After the AGOCS tool modifications, its features were extended to generate datasets in various formats simultaneously for use in ML frameworks. This allowed for rapid testing and comparison of multiple methods. Preparing work traces for simulation is time-consuming, and this research focused on four computing cell traces: clusterdata-2011, clusterdata-2019a, clusterdata-2019c, and clusterdata-2019d.

*B. Constraint Operators Dataset*

This research extended the prior investigation [27] and tested a number of approaches to generate CO datasets. In the course of the investigation, two separate datasets were created from each cell work trace (Figure 1): the COs as Encoded Labels Dataset (CO-EL) and the COs as Value Vectors Dataset (CO-VV), described in detail in the sections below.

Fig. 1. Generation of experimental datasets with AGOCS

*C. Constraint Operators as Encoded Labels Dataset (CO-EL)*

The original method was used, in which the COs are first collapsed (Table V) and used as labels. The result is then One-Hot encoded into a sparse dataset (Table VI), where a given cell has a value of one if the corresponding CO is defined for a task. The main disadvantage of this solution is that newly appearing CO need to be label re-encoded for the given attribute, and as such the model might need to be fully re-trained.

TABLE V. SAMPLE CO COMPACTIONS

| Input CO | Collapsed CO |
|---|---|
| 8 > ${AM}$<br>3 > ${AM}$<br>${AM} > 0 | 3 > ${AM} > 0<br>(operators are compacted into a new Between operator, note that constraint operator 8 > ${AM}$ is obsolete with 3 > ${AM}$ present) |
| ${AM} <> 1<br>${AM} > 3<br>${AM} <> 4 | ${AM} > 4<br>(operators are compacted into a new Between operator, note that the GCD trances support only integer numbers in constrain operators) |
| ${N} <> 'a'<br>${N} <> 'b'<br>${N} <> 'c' | ${N} <> 'a'; 'b'; 'c'<br>(operators are compacted into a new Non-Equal-Array operator) |
| ${G} <> 'a'<br>${G} <> 'b'<br>${G} = 'c' | ${G} = 'c'<br>(Not-Equal operators are removed as Equals operator is restrictive) |
| ${DC} = 1<br>${DC} = 7 | Whenever collapsing COs is not possible, an error will be logged. Such anomalies are very rare (fewer than twenty across all datasets) and are ignored in the simulation as they do not meet the criteria. |

TABLE VI. SAMPLE OF THE CO-EL DATASET (CLUSTERDATA-2011)

[Table of sparse one-hot encoded data with columns: AVAILABLE NODES COUNT, GROUP TASK ID, ${AG} <> 0, ${AH} > 1, ${AH} = 2, ${AK} <> 'hk', ${AM} > 2, ${AM} = 2, ${AN} = 2, ${AS} = =, ${AV} <> 0, ${A} <> 0, ${A} = 2, ${D} = =, ${E} > 3, ${E} > 0, ${E} = 0, 3 > ${E} > 0, ${G} <> 0.1, ${G} <> 0.3, ${G} = 2, ${I} = 4, ${I} > 1, ${N} <> 'ho'; 'hp', ${R} > 0, ${V} > 1, ${W} < 4, ${W} < 5, ${W} > 4, ${W} > 2, 14 > ${W} > 4]

## D. Constraint Operators as Value Vectors Datasets (CO-VV)

In this approach, all possible values for an attribute are listed, with '0' marking acceptable values and '1' marking non-acceptable ones, reversing the common notation since the model focuses on detecting unacceptable nodes. Table VII illustrates this with sample COs for an attribute 'AM'.

TABLE VII. THE REVERSED '0/1' NOTATION OF CO AND MATCHED ATTRIBUTE VALUES

| CO | Attribute 'AM' values vector | | | | | | | | | | |
|---|---|---|---|---|---|---|---|---|---|---|---|
| | ${AM}: (none) | ${AM}:0 | ${AM}:1 | ${AM}:2 | ${AM}:3 | ${AM}:4 | ${AM}:5 | ${AM}:6 | ${AM}:7 | ${AM}:8 | ${AM}:9 |
| ${AM} >= 5 | 1 | 1 | 1 | 1 | 1 | 1 | 0 | 0 | 0 | 0 | 0 |
| 3 > ${AM} > 0 | 1 | 1 | 0 | 0 | 1 | 1 | 1 | 1 | 1 | 1 | 1 |
| ${AM} <> 0; 7; 8 | 0 | 1 | 0 | 0 | 0 | 0 | 0 | 0 | 1 | 1 | 0 |
| ${AM} > 0 | 1 | 1 | 0 | 0 | 0 | 0 | 0 | 0 | 0 | 0 | 0 |

The creation of the values vector allows for the dynamic addition of new features; i.e., another column is appended to the end of the feature array. The main advantage of this method is that the dataset can be extended while the cluster is being reconfigured – additional attributes and their values can be added during cluster operation, and the existing model can be expanded with new input features through TL. Table VIII illustrates the final result.

TABLE VIII. SAMPLE OF THE CO-VV DATASET (CLUSTERDATA-2019A)

[Table VIII: wide dataset sample table with columns including AVAILABLE NODES COUNT, GROUP TASK ID, ${AH}(none), ${AH}:1, ${BC}(none), ${BC}:1, ${BC}:2, ${BU}(none), ${BU}:2, ${BU}:3, ${BU}:5, ${BU}:6, ${BY}(none), ${BY}:1, ${BY}:2, ${BY}:3, ${BY}:4, ${CG}(none), ${CG}:10, ${CG}:11, ${CG}:12, ${CG}:13, ${CG}:2, ${CG}:3, ${CG}:4, ${CG}:5, ${CG}:6, ${CG}:7, ${CG}:8, ${CG}:9, ${CH}(none), ${CH}:1, ${CH}:2, ${CH}:5 — showing various 0/1 values across sample rows]

## E. Task Grouping

A significant portion of tasks in the cluster include CO as part of their scheduling parameters. Table IX presents the distribution of tasks with CO based on volume, requested CPU, and memory ratios across the examined workload trace repositories. Tasks with CO across all analyzed GCD repositories requested an average of 15.9% to 38.2% of CPU and 14.9% to 48.5% of memory, with occasional spikes reaching up to 64.8% of CPU and 74.7% of memory.

TABLE IX. DISTRIBUTION OF TASKS WITH CO BY VOLUME, REQUESTED CPU AND MEMORY

| GCD archive | Tasks with CO by volume | | | Tasks with CO by requested CPU cores | | | Tasks with CO by requested memory | | |
|---|---|---|---|---|---|---|---|---|---|
| | *Min* | *Max* | *Avg.* | *Min* | *Max* | *Avg.* | *Min* | *Max* | *Avg.* |
| clusterdata-2011 | 8.1% | 41.3% | 20.5% | 17.8% | 45.5% | 25.6% | 6.0% | 36.3% | 21.7% |
| clusterdata-2019a | 16.6% | 62.6% | 41.8% | 17.4% | 64.8% | 38.3% | 19.9% | 74.7% | 48.5% |
| clusterdata-2019c | 11.3% | 49.3% | 22.0% | 10.6% | 60.2% | 21.9% | 10.6% | 60.1% | 22.9% |
| clusterdata-2019d | 8.2% | 33.9% | 13.6% | 8.7% | 33.7% | 15.9% | 7.9% | 50.7% | 14.9% |

Tasks with OC occasionally account for over half of the cluster's resources, making it crucial to consider both resource allocation and node-affinity. A key challenge in previous research [26][27] was scheduling tasks with restrictive constraints, where 10-15 tasks per 10,000 required execution on a small subset of nodes, sometimes just one. The 'forced-migration' flag in the negotiation protocol [26] helped but was inefficient, causing workload spikes and premature offloading. Kubernetes uses a similar mechanism where preemption logic evicts lower-priority pods to make room for higher-priority ones. However, Kubernetes' preemption is limited and may block scheduling if no node satisfies affinity rules, leading to errors. Currently, there are no effective solutions for actively monitoring Kubernetes events for allocation failures.

After regression experiments [27], it was found that evaluating ML model performance is easier when tasks are grouped based on the number of suitable nodes. Previous evaluations focused on two groups: tasks allocated to a single node and those requiring 501-1000 nodes, as no tasks in the GCD repository's clusterdata-2011 required 2-500 nodes. This research considered the former group unnecessary, as a few hundred nodes suffice for smooth cluster scheduling. The focus was on tasks that can run on a single node and overall model accuracy. Tasks are divided into 26 groups, with Group 0 for tasks allocated to a single node and Groups 1–25 based on increments of 500 suitable nodes. For clusterdata-2019a, tasks are grouped every 360 nodes due to its smaller cell size (9.4k nodes). These 26 groups are used to train and test classifying ML models.

## IV. DYNAMICALLY GROWING MODEL

TL is an ML technique where a model trained on one task is adapted to a related task, using knowledge from pre-trained models instead of starting from scratch. In this research, the CO-VV model was designed to be extensible during cluster operations, allowing the previously created ML model to accommodate new values for node attributes. Table XI shows how the feature array grew over thirty-one days of simulation for a sample computing cell clusterdata-2019c, with most attribute values defined in step zero. For traceability and simplicity, new attribute values are appended as the last column.

The dataset features are frequently extended, requiring the model to be adjusted accordingly. To handle dynamic changes in model weight structures, the project [27] transitioned from SciKit-learn to PyTorch. This shift was necessary to leverage PyTorch's advanced capabilities for adaptive data structures. The framework offers a flexible deep learning framework for defining custom NN, optimizing gradients, and managing computations at a granular level. It enables direct manipulation of tensors, making it ideal for tasks requiring dynamic model reconfiguration. In contrast, SciKit-learn prioritizes simplicity and ease of use, offering limited customization mainly through hyperparameter tuning. While SciKit-learn excels in tasks like test data splitting and cross-validation, PyTorch provides greater flexibility for evolving datasets.

Despite the switch to PyTorch, SciKit-learn remains valuable for pre-processing, evaluation, train/test datasets splits, and baseline comparisons. The hybrid approach combines

PyTorch for dynamic model adaptation and SciKit-learn for utility functions and prebuilt algorithms, optimizing workflows. This integration also links efficient data manipulation (Pandas and NumPy) with ML algorithms (SciKit-learn). Pandas and NumPy simplify data cleaning, transformation, and analysis. The following package versions were used: Python 3.12.2, SciKit-learn 1.5.1, NumPy 2.1.2, Pandas 2.2.3, and PyTorch 2.6.0.dev20241126.

The model training process is complex, with the same code responsible for initializing new models. Figure 2 illustrates the training stages, followed by key routines.

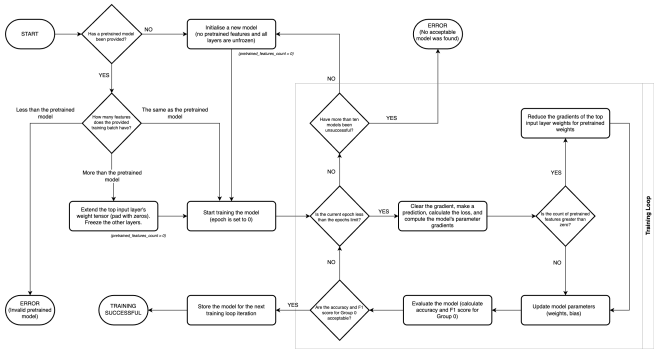

Fig. 2. The CTL-based Incrementally Expanding model's training routine

### A. Extending Input Layers

In traditional ML, a new model would need to be created with an input layer matching the updated feature size, followed by full retraining. To optimize, TL can be used to transfer the learned state to the new model. This is more effective in multi-layer models, where deeper layers capture generalized knowledge and are frozen, while the top layers are fine-tuned. In PyTorch, this is done by setting the 'requires_grad' attribute of layer parameter tensors to False. After restoring a trained model, base layers are frozen.

```
# create two-layer model, 30 neurons in the hidden layer
model = nn.Sequential(OrderedDict([
    ('fc1', nn.Linear(dataset_data.features_count, HIDDEN_LAYER_SIZE)),
    ('fc2', nn.Linear(HIDDEN_LAYER_SIZE, CLASSES_COUNT))  # Groups 0-25
]))
model = model.to(device=device, dtype=torch.float32)

# restore model from a file
model_state_dict = torch.load(model_file_path)
model.load_state_dict(model_state_dict)

# freeze all base layers
for param in model.features.parameters():
    param.requires_grad = False
```
Listing 1. Loading the model's saved state

Prior research [27] showed that for highly sparse data, where ones represent less than 0.01% of the total, a multi-layer model is unnecessary. A three-layer model achieved 98% accuracy. Therefore, this research focused on modifying the top input layer without losing acquired knowledge. New values are appended to the end of the features array, and only input weights are extended, while the number of neurons in the layer remains unchanged.

Listing 2 outlines the process of extending the input layer. The reshaping occurs within the model's state dictionary before restoring the model, simplifying the code by automatically reinitializing all internal values. Since the CO-VV dataset appends new values to the end of the features array, initializing the new weights to zero ensures compatibility with the previous dataset, where new attribute values do not exist yet.

```
# restore model from a file
model_state_dict = torch.load(model_file_path)
model.load_state_dict(model_state_dict)

fc1_weight_tensor = model_state_dict['fc1.weight']
pretrained_features_count = fc1_weight_tensor.size(dim=1)

# extend input layer's weights
if pretrained_features_count != dataset_data.features_count:
    fc1_weight_tensor = torch.nn.functional.pad(
        input=fc1_weight_tensor,
        # padding on the right side
        pad=(0, dataset_data.features_count - pretrained_features_count),
        mode='constant',
        value=0  # pad with zeros
    )
    # replace parameter tensor in model dict
    model_state_dict['fc1.weight'] = fc1_weight_tensor

# restore model
model.load_state_dict(model_state_dict)
```
Listing 2. Extending the model's top input layer weights

### B. Dynamic Gradient Modifications

The modified model retains knowledge from previous training and can predict classes in new datasets but requires retraining due to additional features. However, traditional TL was ineffective for models with extended input layers, resulting in suboptimal performance. To address this, the old weights were minimally altered, while the new weights (introduced by padding the 'fc1.weight' tensor) were trained more extensively. Listing 3 shows the training loop, which dynamically adjusts the gradient tensor for the top input layer's previously trained weights.

```
# create weighted loss function (assign higher weight to Group 0)
class_weights = torch.tensor(
    data=[GROUP_0_CLASS_WEIGHT] + [1] * 25,
    dtype=torch.float)
loss_function = torch.nn.CrossEntropyLoss(weight=class_weights)

# create Adam optimizer with learning rate of 0.05
optimizer = torch.optim.Adam(model.parameters(), lr=0.05)

# create multiplier tensor in device memory:
# [0.1, 0.1, 0.1, ... 1, 1]
multiplier_tensor = torch.FloatTensor(
    [PRETRAINED_GRADIENT_RATE] * pretrained_features_count +
    [1] * (dataset_data.features_count - pretrained_features_count),
    device=device
)
multiplier_tensor.requires_grad=False

# training loop
for epoch in range(EPOCHS_LIMIT):
    model.train()  # set train mode
    for X_batch, y_batch in dataset_data.train_loader:

        # clear gradient, make prediction, calculate logits and loss
        optimizer.zero_grad()
        y_logits = model(X_batch)
        loss = loss_function(y_logits, y_batch)

        # calculate gradients of parameters
        loss.backward()

        for name, param in model.named_parameters():
            if name == 'fc1.weight':
                # multiply gradient tensors in fc1 layer's weights
                with torch.no_grad():
                    for index, param_grad in enumerate(param.grad):
                        # in-place multiplication
                        param_grad.mul_(multiplier_tensor)
                # enable weights for training
                param.requires_grad = True
            elif name == 'fc1.bias':
                # enable bias for training
                param.requires_grad = True
            else:
                # other layers are frozen
                param.requires_grad = False

        # update model parameters
        optimizer.step()

    # evaluate model
    model.eval()
    accuracy, group_0_f1_score = evaluate_model(dataset_data.X_test,
                                                 dataset_data.y_test,
                                                 model)
    # early stop when accuracy and f1 score are acceptable
    if (
        accuracy > ACCEPTED_ACCURACY and
        group_0_f1_score > ACCEPTED_GROUP_0_F1_SCORE
    ):
        # exit training loop
        break
```
Listing 3. The growing model training loop

The Adam optimizer (torch.optim.Adam) with a learning rate of 0.05 and the Cross-Entropy loss function (torch.nn.CrossEntropyLoss) were used. Adam, based on RMSProp, adjusts the step size over time and incorporates momentum, making it suitable for sparse gradients or noisy data. While it converges faster than SGD, it may not generalize as well. Cross-Entropy loss is ideal for classification tasks, heavily penalizing incorrect predictions. In this model, the class weight for Group 0 was increased by 200 (group_0_class_weight) to prioritize accurate classification of tasks allocable to a single node. Backpropagation computes gradients using the chain rule, and in the modified training loop, the gradient tensors for pre-trained weights are scaled by a factor of 0.1 (pretrained_gradient_rate) to reduce their learning rate, while newly added weights retain their original gradients. Through experimentation (which also helped set other values), it was found that a scaling factor above 20-30% negated training effects, while zeroing gradients for pre-trained weights reduced model accuracy.

PyTorch Autograd computes gradients by reversing the computation graph using the chain rule, supporting scalar and tensor operations for multi-dimensional data. The graph is dynamically built during the forward pass [28], enhancing flexibility and memory efficiency. To optimize performance, the multiplier tensor is created once, loaded into device memory, and used with an in-place function (torch.Tensor.mul_). Operations are performed within a torch.no_grad block to prevent unnecessary Autograd recording. The training loop includes an early exit mechanism, terminating when accuracy exceeds 0.95 (accepted_accuracy) and the F1 score for Group 0 exceeds 0.9 (accepted_group_0_f1_score). These limits were derived from the baseline results reported in [27]. If these thresholds are not met within 100 epochs (epochs_limit), the pre-trained model is discarded, and a new one is initialized, ensuring a fail-fast approach. Training halts after ten failed attempts to prevent excessive resource use.

## V. Model Evaluation

The proposed **Growing** model was compared to a **Fully Retrain** variant, which fully retrains on each step's dataset, and baseline SciKit-learn models known for handling large, sparse datasets efficiently [27][32]. The baselines included:

- **MLP Classifier** (sklearn.neural_network.MLPClassifier): Delivered strong results with default hyperparameters, further improved through tuning. Similar to the Growing model, the ANN was configured with 30 hidden units and the default Adam optimizer.

- **Ridge Classifier** (sklearn.linear_model.RidgeClassifier): Uses Ridge Regression, which adds an L2 regularization penalty to prevent overfitting by discouraging large coefficients. It is computationally efficient, interpretable, and effective for datasets with many features or correlated variables.

- **SGD Classifier** (sklearn.linear_model.SGDClassifier): Implements a Linear SVM trained with Stochastic Gradient Descent, optimizing weights incrementally for each data point. This approach is fast, memory-efficient, and suitable for high-dimensional problems like text classification.

- **Ensemble Voter** (sklearn.ensemble.VotingClassifier): Combines predictions from the baseline models using hard voting, as some models lacked the 'predict_proba' method needed for soft voting.

The four computing cells: clusterdata-2011, clusterdata-2019a, clusterdata-2019c, and clusterdata-2019d, were evaluated individually. Stratified training and testing datasets were created where possible (at least two samples per class were required). The evaluation focuses on overall accuracy and Group 0 performance, with Group 0 tasks making up only 0.03% to 1.17% of total tasks, reflecting significant class imbalance. Stratified randomized folds were used to preserve class proportions, ensuring balanced representation despite the computational cost. Table X summarizes the results for accuracy and Group 0 F1 scores, while Table XI presents a detailed run using clusterdata-2019c as a sample. Each step marks the simulation time (day, hour, minute) when the feature array was extended, prompting model retraining. Except for the Growing model, all models were trained from scratch, with epoch counts noted for ANN models. Group 0 F1 scores are omitted when no Group 0 samples were present in the test dataset.

The evaluation routines were run on a 2023 MacBook Pro (Apple M2 Pro, 12-core CPU, 16GB RAM). Timings were measured for instances where the model from the previous iteration needed retraining and included all necessary steps, including model loading, dataset splitting (with stratification), training, and evaluation. For baseline models, the MLP Classifier took 7-29 minutes per step, Ridge Classifier 11-23 minutes, SGD Classifier 12-37 minutes, and Ensemble Voter (which is well-parallelized) took 19–42 minutes. For the developed models, the Fully Retrained version took 8-33 minutes (similar to the MLP Classifier), while the Growing model took 17 minutes for the initial model training and 1-6 minutes for each subsequent step.

A key observation is the higher accuracy and improved Group 0 F1 scores for all models compared to 2019 datasets, attributed to efficiency improvements in Google's Borg between 2011 and 2019, including alloc and parent-child dependencies. During this period, task submission rates increased 3.7-fold, total tasks grew sevenfold, and scheduling time remained stable. Task resource consumption exhibited heavy-tailed Pareto distributions, with the top 1% of tasks consuming over 99% of total resources. These changes made workload traces more challenging and model building more complex [3].

TABLE X. SUMMARY OF MODEL EVALUATION RESULTS

| Dataset | Model | | | | | | | | | | | | | |
|---|---|---|---|---|---|---|---|---|---|---|---|---|---|---|
| | Growing | | | Fully Retrain | | | MLP Classifier | | Ridge Classifier | | SGD Classifier | | Ensemble Voter | |
| | Avg. accuracy | Avg. Group 0 F1 score | Epochs total | Avg. accuracy | Avg. Group 0 F1 score | Epochs total | Avg. accuracy | Avg. Group 0 F1 score | Avg. accuracy | Avg. Group 0 F1 score | Avg. accuracy | Avg. Group 0 F1 score | Avg. accuracy | Avg. Group 0 F1 score |
| clusterdata-2011 | 0.99957 | 1 | 66 | 0.9988 | 0.9988 | 746 | 0.98676 | 0.99509 | 0.9989 | 1 | 0.9921 | 1 | 0.99949 | 0.97107 |
| clusterdata-2019a | 0.9918 | 1 | 107 | 0.98823 | 0.99824 | 179 | 0.98007 | 0.63636 | 0.98182 | 0.7953 | 0.98206 | 0.64182 | 0.98264 | 0.58333 |
| clusterdata-2019c | 0.98581 | 0.99919 | 76 | 0.98844 | 0.95834 | 830 | 0.88443 | 0.89052 | 0.99382 | 0.97252 | 0.98857 | 0.90708 | 0.99419 | 0.93742 |
| clusterdata-2019d | 0.99416 | 1 | 161 | 0.98774 | 0.99615 | 261 | 0.93616 | 0.91278 | 0.99789 | 0.89427 | 0.99896 | 0.80121 | 0.99844 | 0.98899 |

TABLE XI. MODEL EVALUATION RESULTS FOR CLUSTERDATA-2019C

All models performed well, with minimal differences in accuracy and Group 0 F1 scores. While baseline classifiers (MLP, Ridge, and SGD) were slightly less consistent, incorporating them into a hard-voting ensemble produced reliable results [27]. Both researched model variants Growing and Fully Retrain also delivered satisfactory performance, highlighting the capabilities of modern ML frameworks like PyTorch and SciKit-learn. However, baseline models lack flexibility and the ability to retain learned knowledge, limiting their adaptability compared to ANN-based designs. The Growing model's key advantage is its computational efficiency, requiring significantly fewer epochs for training: 40% fewer for clusterdata-2019a and up to 91% fewer for clusterdata-2019c. This efficiency stems from reusing pre-trained layers' weights and biases, drastically reducing training time. While baseline models (MLP, Ridge, SGD, and Ensemble Voter) achieved good results, their training times were at least an order of magnitude longer. In contrast, the Growing model operates almost in real time, enabling rapid evaluation of cluster task queues as tasks arrive. This opens opportunities for specialized schedulers to optimize task allocation more effectively.

## VI. Conclusions

The key contribution of the research is the introduction of the CTLM, which helps detect and prioritize tasks with restrictive node-affinity constraints. Compared to previous research [27], the models improved accuracy from 98% to 99% and achieved higher F1 scores for Group 0. This improvement is largely due to a shift in data encoding from the CO-EL format (COs are one-hot encoded as labels) to the CO-VV format (using node attribute values directly as labels). While the feature array size increased substantially from 4.4k to ~16k, both baseline and newly introduced ANN-based models handled the larger arrays effectively.

The adoption of the CO-VV model introduced the challenge of updating trained models to accommodate new features by extending model inputs as values were added. While fully retraining the model after each extension is feasible, the research prioritized designing a dynamically growing model that retains prior knowledge. Given the simplicity of the two-layer ANN, the typical TL approach, i.e., removing and freezing top layers, was considered suboptimal. Instead, the research focused on dynamically extending the top input layer by adding new weights.

To implement and evaluate the dynamically extended model, the project transitioned from SciKit-learn [27] to PyTorch for finer control over training and direct manipulation of model data. Despite this shift, the codebase retained routines from both frameworks, with evaluations referencing SciKit-learn baseline models. PyTorch's flexibility enabled the implementation, though it required low-level coding, including direct modifications to the model state dictionary and dynamic padding of layer parameter tensors. The resulting growing model achieved comparable accuracy and Group 0 F1 scores to baseline and fully recreated models while requiring 40% to 91% fewer epochs for retraining.

The reduction in training epochs, along with improved accuracy, makes the dynamically growing model ideal for near real-time applications. It can enhance cluster orchestration systems by rerouting high-priority tasks to specialized allocation strategies before the main cluster scheduler processes the pending job queue, as shown in Figure 3. This approach works well with gang scheduling, where tasks in the same job are grouped by their CO and scheduled together. Coordinating with the Main Cluster Scheduler, the High-Priority Scheduler minimizes task scheduling latency by prioritizing tasks with fewer suitable nodes. Additionally, updating ML model runs in parallel and won't block or slow down the main cluster scheduler.

The presented schema was tested using real-world GDC traces; however, it is not suitable for every scenario. During the investigation, the following shortcomings were noted:

- Adding new features to the ANN should be done gradually. Experimentation showed that adding over 40–50 features at once often reduces accuracy and forces full model retraining.

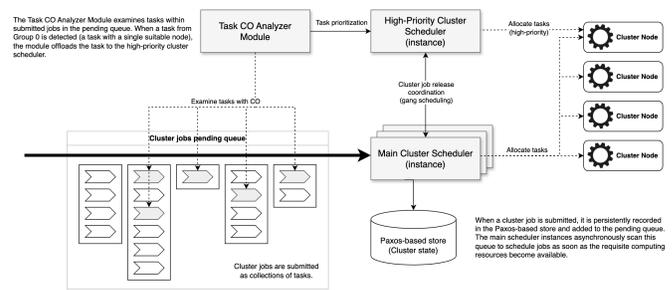

Fig. 3. Enhanced Cluster Job Scheduling with the Task CO Analyzer Module

- The growing model approach worked well for the CO-VV dataset but not for CO-EL, as CO-VV features can be grouped for generalization, while CO-EL's label-encoded COs lack overlapping properties for effective generalization.

- The codebase required low-level PyTorch routines, such as in-device, in-place weight multiplication, a feature not available in other frameworks like TensorFlow and MXNet.

The research objectives were successfully met, with improved accuracy and F1 scores. The model now incorporates new constraints in near real-time without full retraining, making it suitable for ongoing use. Additionally, transitioning to PyTorch supports future research and iterations. The investigation has revealed several promising directions for future research:

- **Task Misclassification via Hybridization**: A mixed model that combines ML with predefined rules (human input). Misclassifying single-node tasks as multi-node ones, while manageable, may cause performance issues like resource reallocation. A secondary heuristic layer could better handle edge cases, reducing disruptions.

- **Expiring Unused Attributes**: While this wasn't an issue in the thirty-one-day simulation, more active cluster configurations may face challenges if unused attribute values accumulate over time. Introducing a process to retire obsolete features will keep the model efficient and scalable.

- **Enhancing Scheduler Algorithms**: The proof-of-concept models can be applied to existing schedulers, with Kubernetes as a promising candidate for further experimentation in real-world scenarios.

- **Broadening Work Trace Analysis**: Without extensive testing, it's uncertain whether this method works on a larger scale. Experiments using work traces from other sources like the Open Grid Workload Archive or supercomputers could improve research, but access challenges remain due to incomplete or restricted data.

- **Investigating Node 'Soft' Affinity**: Kubernetes' 'soft' node-affinity adds complexity to scheduling, necessitating further research to optimize its application in cluster management.

ACKNOWLEDGMENT

The authors of this paper would like to thank Google engineers, and in particular John Wilkes, for describing the internal workings of the Borg scheduler and enabling access to detailed Google Cluster Data workload traces.